\documentclass[amssymb,twocolumn,aps]{revtex4}
\usepackage{graphics,color,graphicx,amsmath}
\usepackage{subcaption}
\usepackage{verbatim}
\usepackage{graphicx}
\usepackage[above,below]{placeins}
\usepackage{float}
\usepackage{tabularx}
\usepackage{times}
\usepackage{blindtext}
\usepackage{url}
\usepackage{rotating}
\usepackage{mathtools}

\begin{document}

\title{Relative benefits of different active learning methods to conceptual physics learning}


\author{Meagan Sundstrom~\footnote[1]{sundstrommeagan@gmail.com}$^{1}$, Justin Gambrell$^{2}$, Colin Green$^{3}$, Adrienne L. Traxler$^{4}$, and Eric Brewe~\footnote[2]{Corresponding author: eb573@drexel.edu}$^1$}
\affiliation{$^1$Department of Physics, Drexel University, Philadelphia, Pennsylvania 19104, USA\\
$^{2}$Department of Computational Mathematics, Science and Engineering, Michigan State University, East Lansing, Michigan 48824, USA\\
$^3$Department of Physics, Bryn Mawr College, Bryn Mawr, Pennsylvania 19010, USA\\
$^4$Department of Science Education, University of Copenhagen, Copenhagen, Denmark}

\date{\today}

\begin{abstract}
It has been shown that active learning methods are more effective than traditional lecturing at improving student conceptual understanding and reducing failure rates in undergraduate physics courses. Researchers have developed distinct, active learning methods that are now widely implemented in introductory physics. However, the relative benefits of these methods remain unknown. Here we present a multi-institutional comparison of the impacts of four well-established active learning methods -- Peer Instruction, Investigative Science Learning Environment (ISLE), Tutorials and Student-Centered Active Learning Environment with Upside-Down Pedagogies (SCALE-UP) -- on conceptual learning. We find measurable increases in student conceptual learning in all four active learning methods, and significantly larger gains in SCALE-UP than in either Peer Instruction or ISLE. Student development of peer networks is similar across the four methods, but classroom activities differ. In many of the observed Peer Instruction and ISLE courses, instructors lecture for a large fraction of class time. In Tutorials and SCALE-UP courses, instructors dedicate most in-class time to student-centred activities such as worksheets and laboratory work. These results prompt future work to identify causal mechanisms between specific classroom activities and conceptual learning and to examine additional factors related to variation in student learning across different methods.

\end{abstract}

\maketitle

Active learning invites students to cognitively engage with course material during class time and often involves students interacting with their peers such that they construct their understanding together~\cite{lombardi2021curious}. Active learning strategies are demonstrably better at promoting conceptual development and improving odds of student success than traditional, lecture-based teaching methods in university-level physics courses ~\cite{hake1998interactive,freeman2014active,theobald2020active}.  It is no longer relevant to compare active learning to traditional methods~\cite{wieman2014large,dancy2024physics}; rather, researchers must measure the ``relative benefits of different active learning methods"~\cite[p. 8320]{wieman2014large} and ``better [understand] the strengths and weaknesses of different types of active learning instructional styles"~\cite[p. 9]{dancy2024physics}. 

While active learning refers to a wide range of research-based instructional strategies, including specific activity types (e.g., computer simulations), it is often discussed and studied through the names of well-established methods. Researchers have started to compare these named active learning methods, but they have only done so at single institutions and/or for high-fidelity implementations of the methods (e.g., at the sites where the methods were developed)~\cite{rogers2015implementing,keiner2010interactive,commeford2021characterizing,commeford2022characterizing,weir2019small}. As a growing number of instructors implement these methods in their physics courses~\cite{dancy2024physics,foote2014diffusion}, it is necessary to understand the relative benefits of each method to conceptual learning, and the mechanisms that explain these benefits, across a variety of implementations.

Here we report the first multi-institutional and multi-dimensional study of the relative impacts of named active learning methods on introductory physics and astronomy students' conceptual learning. We examine 31 total implementations of four methods:
\begin{enumerate}
    \itemsep0cm
    \item Peer Instruction~\cite{mazur1997peer}: During lectures, students work in small groups of nearby peers to answer clicker (or other voting system) questions. Typically, the instructor poses a question, students answer the question individually, students discuss the question in small groups, students re-answer the question, and then the instructor explains the answer. 
    \item Investigative Science Learning Environment (ISLE)~\cite{etkina2007investigative}: In all course components, or only in laboratories (labs), students engage in scientific processes in small groups. Students observe a physics experiment, explain their observations, make predictions about new experiments, design and conduct these experiments, and revise their explanations.
    \item Tutorials in Introductory Physics~\cite{mcdermott2002tutorials} and Astronomy~\cite{adams2003lecture}: During lecture and/or recitation sections, students complete worksheets in small groups. These worksheets intend to elicit, confront, and resolve common misconceptions. 
    \item Student-Centered Active Learning Environment with Upside-down Pedagogies (SCALE-UP)~\cite{beichner2007student}: Students solve problems and complete laboratory activities in small groups in an integrated learning (or studio-style) environment, often containing large tables that seat nine students and whiteboards along the classroom perimeter.
\end{enumerate}
We also investigate two central features of active learning that may explain differences between each method's impact on student conceptual learning: the development of peer networks and the amount of in-class time spent on student-centered activities. 

We examine peer networks because student-student collaboration is a defining feature of active learning and several research studies have found that students who engage in more interactions with their peers tend to earn higher grades in undergraduate physics courses~\cite{bruun2013talking,williams2019linking,sundstrom2022interactions}. At the same time, prior work indicates that the structures of peer networks may vary between different active learning methods. In one implementation of Peer Instruction and one implementation of SCALE-UP, for example, researchers observed that students formed string-like patterns of connections to peers during Peer Instruction (mimicking rows of chairs in a lecture hall) and well-connected small groups during SCALE-UP (mimicking tables set up for groupwork)~\cite{commeford2021characterizing}. Some active learning methods, therefore, may improve student conceptual understanding more than others if they foster certain types of peer networks (e.g., having a small group of frequently interacting peers might be more beneficial to conceptual learning than having sparse interactions with nearby peers). Here we statistically model peer network dynamics from the beginning to the end of the semester across four active learning methods to better understand this possibility. 

We examine classroom activities because prior studies have found an overall positive correlation between the extent of active learning used (e.g., percentage of class time spent on lecturing versus active learning) and student understanding~\cite{rogers2015implementing,keiner2010interactive,prather2009national}. Previous research has also identified that the names of active learning methods may not clearly map onto instructional practices~\cite{vishnubhotla2024use,turpen2009not,wood2016characterizing,commeford2022characterizing}.
One study, for example, observed that even among instructors at the same institution who all used Peer Instruction, the amount of time given to students to answer the clicker questions and the amount of time the instructor spent explaining the answers varied widely~\cite{turpen2009not}. The latter studies, however, do not link specific instructional activities to conceptual learning -- they only measure and characterize the frequencies of the activities. Yet, it is plausible that there are relative benefits of particular activities within an active learning method to student conceptual understanding~\cite{weir2019small}. We empirically characterize classroom activities (i.e., rather than relying on the method names to infer classroom activities) and compare them to student conceptual learning for the four active learning methods studied here to investigate this relationship.

We recruited the instructors of 31 introductory physics and astronomy courses at 28 different institutions to participate in the study. All courses were first-semester courses (e.g., mechanics) in which the instructor self-reported using Peer Instruction, ISLE, Tutorials, or SCALE-UP.  The institutions include both public and private universities and a range of research-intensive and undergraduate-focused universities (Supplementary Information Table 2). Institutions also span most geographic regions of the continental United States. Three institutions are Hispanic-serving and one institution is a women's college.

We asked each instructor to collect three sources of data: (1) pre- and post-semester concept inventories, (2) pre- and post-semester network surveys, and (3) video recordings of three consecutive class sessions.

Concept inventories, such as the Force Concept Inventory~\cite{hestenes1992force}, are research-validated instruments used to measure student understanding of physics content. Instructors chose an existing concept inventory to give to their students, such that the concept inventory aligned with the topics taught in the course, and administered the assessment online (e.g., using the Learning About STEM Student Outcomes tool~\cite{lasso}) at both the beginning (before any exposure to the material) and the end of the semester (Supplementary Information Table 3). 26 of the 31 courses had at least 40\% of enrolled students with matched responses (i.e., completing both the pre- and post-semester concept inventory) and/or at least 30 enrolled students with matched responses and are included in our analysis of conceptual learning (Supplementary Information Table 1). 

Instructors also administered an online network survey using Qualtrics at both the beginning and the end of the semester. The survey probed which peers students interacted with using the following prompt adopted from prior work~\cite{commeford2021characterizing}: ``Please choose from the list of people that are enrolled in your physics class the names of any other student with whom you had a meaningful interaction in class during the past week, even if you were not the main person speaking." The prompt was followed by a list of all students enrolled in the course with associated checkboxes. Students could select as many of their peers' names as they wanted. 19 of the 31 courses had both pre- and post-semester response rates above 50\% of enrolled students on the survey and are included in the network analysis (Supplementary Information Table 1).

For the classroom video recordings, instructors used Zoom, their own camera, or a camera that we provided to them via mail to record three consecutive class sessions, capturing a ``typical week" in the course~\cite{weir2019small, stains2018anatomy}. All courses were conducted in person and videos included both students and the instructor(s) in the camera's view. 30 of the 31 courses had three full video recordings and are included in our analysis of classroom activities (Supplementary Information Table 1).

\begin{figure}[t]
    \centering
\includegraphics[width=3.4in, trim = {0 0cm 0 0}]{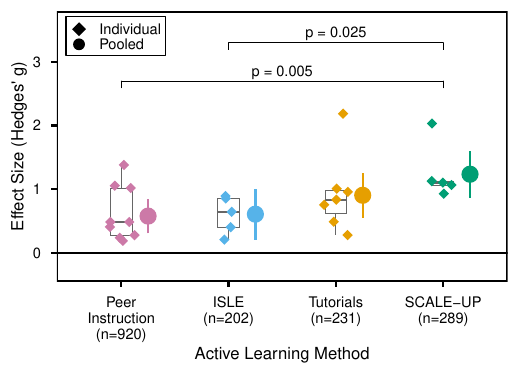}    \caption{\textbf{Students in SCALE-UP courses exhibit significantly larger conceptual learning gains than students in either Peer Instruction or ISLE courses.} Effect sizes for concept inventory scores by active learning method (n = 1,642 students). Diamonds represent Hedges' \textit{g} values for individual courses. Gray boxes indicate the interquartile range, with the lower edge of each box marking the first quartile (25th percentile) and the upper edge marking the third quartile (75th quartile). The bold horizontal line inside each box represents the median (50th percentile). Whiskers extend from the box edges to the most extreme data points within $1.5\times$ the interquartile range. Circles represent Hedges' \textit{g} values pooled by active learning method, with error bars indicating 95\% confidence intervals. Horizontal brackets indicate statistically significant pairwise differences in pooled effect sizes using a mixed effects regression model with effect size as the dependent variable and active learning method as a moderator (see all \textit{p}-values in Supplementary Information Table 4). \textit{N} values indicate the number of students included in the analysis for each method.
    }
   
    \label{fig:main}
\end{figure}

\begin{figure*}[htb]
    \centering
    \includegraphics[width=6in, trim = {0 0cm 0 0}]{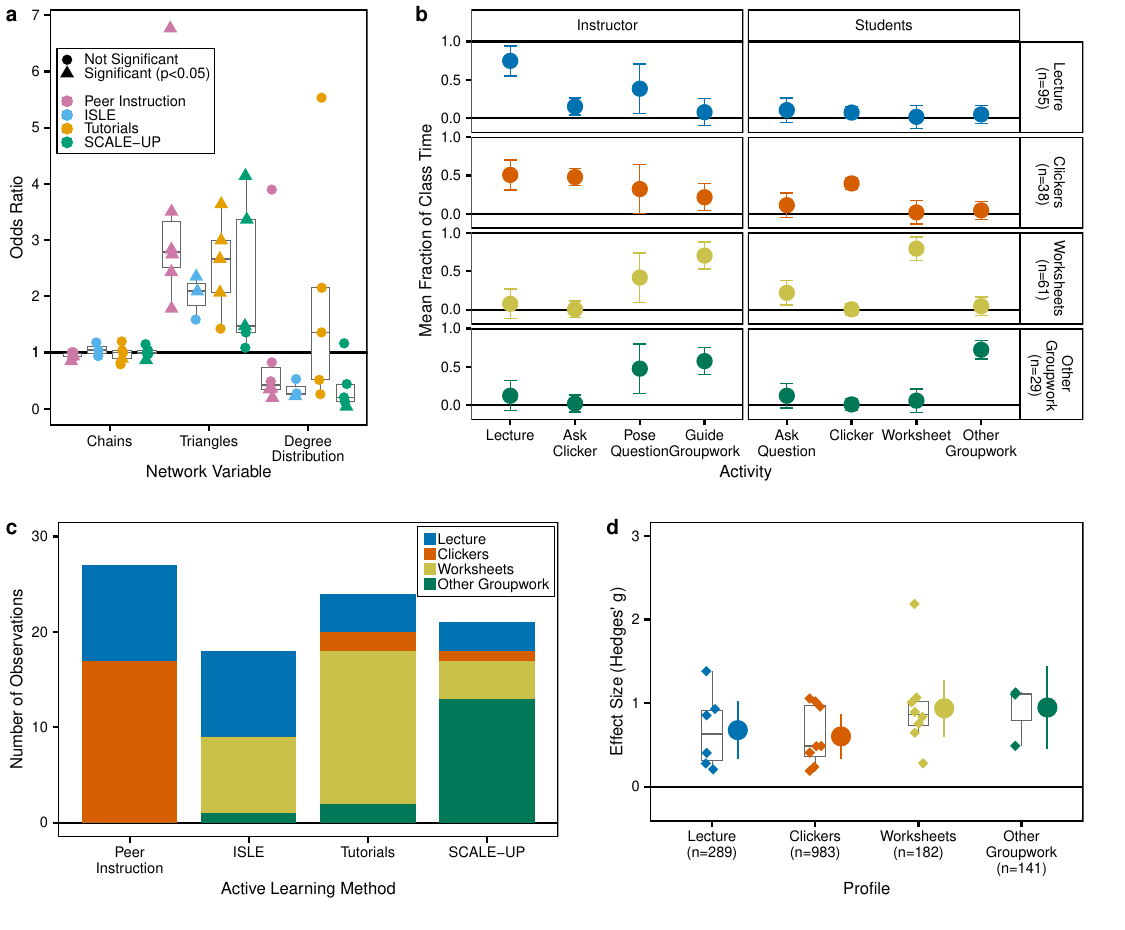}   
    \caption{\textbf{Variation in student conceptual learning across methods is not related to peer networks, but may be related to in-class activities. a}, Results of temporal exponential random graph models applied to peer network data (see variable definitions in Supplementary Information Table 5 and all coefficient values and associated \textit{p}-values in Supplementary Information Table 6; n = 2,033 students in n = 19 courses). Points represent coefficient estimates for each individual course and gray boxes indicate the interquartile range, with the lower edge of each box marking the first quartile (25th percentile) and the upper edge marking the third quartile (75th quartile). The bold horizontal line inside each box represents the median (50th percentile). Whiskers extend from the box edges to the most extreme data points within $1.5\times$ the interquartile range. Odds ratios that are greater (less) than one indicate that the network structure represented by the predictor variable is more (less) likely to form in our observed post-network than we would expect if the pre-network evolved randomly. \textbf{b}, Results of Latent Profile Analysis of classroom observations, including the number of observations assigned to each profile (n = 223 observations). Points indicate mean fractions of class time spent on each activity (calculated as the mean number of two-minute time intervals in which each activity was present) for observations assigned to each profile and error bars indicate standard deviations (see activity definitions in Supplementary Information Table 8 and values in Supplementary Information Table 9).
     \textbf{c}, The number of observations of each active learning method assigned to each latent profile for the data collected in the current study (n = 90 observations).
      \textbf{d}, Effect sizes for concept inventory scores by profile (n = 1,595 students). Diamonds represent Hedges' \textit{g} values for individual courses. Gray boxes indicate the interquartile range, with the lower edge of each box marking the first quartile (25th percentile) and the upper edge marking the third quartile (75th quartile). The bold horizontal line inside each box represents the median (50th percentile). Whiskers extend from the box edges to the most extreme data points within $1.5\times$ the interquartile range. Circles represent Hedges' \textit{g} values pooled by profile, with error bars indicating 95\% confidence intervals. \textit{N} values indicate the number of students included in the analysis for each profile.
    }
   
    \label{fig:main2}
\end{figure*}

\section*{Results}

Similar to prior work~\cite{freeman2014active}, we measured student conceptual learning by calculating an effect size, Hedges' \textit{g}, for student concept inventory scores both within each course and pooled by active learning method~\cite{cohen1992quantitative,gurevitch1999statistical,torchiano2016effsize,viechtbauer2010conducting}. Hedges' \textit{g} can be interpreted as a weighted standardized mean difference between students' pre- and post-semester concept inventory scores. 

All four pooled effect sizes are significantly larger than zero, indicating that student conceptual understanding significantly increased in all four active learning methods (Peer Instruction: Hedges' \textit{g} = 0.58 [95\% confidence interval: 0.31, 0.85], 4.18$\sigma$ difference from a null effect, $p<$0.001; ISLE: Hedges' \textit{g} = 0.61 [0.21, 1.01], 2.97$\sigma$ difference from a null effect, $p$ = 0.003; Tutorials: Hedges' \textit{g} = 0.90 [0.55, 1.26], 4.99$\sigma$ difference from a null effect, $p<$0.001; SCALE-UP: Hedges' \textit{g} = 1.24 [0.86, 1.61], 6.52$\sigma$ difference from a null effect, $p<$0.001; circles in Fig.~\ref{fig:main}). A heterogeneity analysis also indicates statistically significant variation in conceptual learning gains across active learning methods ($Q$ = 9.06, \textit{df} = 3, \textit{p} $=$ 0.029). The pooled effect size for SCALE-UP is significantly larger than the pooled effect sizes for Peer Instruction ($p=$ 0.005) and ISLE ($p=$ 0.025). No other pairs of methods have significantly different pooled effect sizes (Supplementary Information Table 4). Visually, the lower bound of the distribution of individual course effect sizes is substantially higher for SCALE-UP than for the other three methods (diamonds and associated boxplots in Fig.~\ref{fig:main}). 



We investigated two mechanisms that may explain these differences in student conceptual learning across methods: peer networks and classroom activities. To examine the types of peer networks that students develop, we created two undirected networks per course (one pre-semester and one post-semester) with nodes representing students and edges representing reported interactions between students. For each course, we applied a separable temporal exponential random graph model to measure whether different structures were more (or less) likely to form in the post-semester network than we would expect if the pre-semester network evolved randomly~\cite{krivitsky2014separable,hunter2008goodness}. We included four predictor variables in the model: edges (intercept term counting the total number of edges between students), chains (count of unclosed triangles of edges among three students, measuring string-like structures), triangles (count of closed triangles of edges among three students, measuring small group structures), and degree distribution (count of number of edges connected to each student; Supplementary Information Table 5).

Peer network development is comparable across all four active learning methods (signs and significance of most coefficients are similar for each variable in Fig.~\ref{fig:main2}a). In almost all courses, students were not more or less likely to develop string-like chains of connections to peers than we would expect at random (odds ratios close to one for chains), but were significantly likely to develop connections with small groups of peers (odds ratios greater than one for triangles). With the exception of Tutorials courses, many students were likely to form a few connections with peers rather than only a few students forming many connections with peers (i.e., odds ratios less than one for degree distribution). In Tutorials, the odds ratios for the degree distribution variable were not distinguishable from one in any course due to large standard errors on the estimates (Supplementary Information Table 6). The ways that students develop peer networks, therefore, likely do not explain the observed variation in conceptual learning gains (Fig.~\ref{fig:main}).

We then examined possible variation in the activities that took place during instruction within each active learning method. We applied the Classroom Observation Protocol for Undergraduate STEM (COPUS)~\cite{smith2013classroom,landis1977measurement} to 114 h of classroom video collected across 90 observations of 30 courses in our study (three observations per course). The COPUS records the co-occurrence of certain behaviors of the instructor (e.g., lecturing, asking a clicker question, posing a question, and guiding groupwork) and the students (e.g., asking a question, answering a clicker question in groups, completing a worksheet in groups, and completing other forms of groupwork) at two-minute intervals (Supplementary Information Table 8). We combined our 90 COPUS observations with 133 comparable COPUS observations from two prior studies of introductory physics courses (to reach a sufficient sample size for reliable results)~\cite{commeford2022characterizing,stains2018anatomy} in a Latent Profile Analysis to identify groups of similar observations, or ``profiles"~\cite{masyn2013latent-key,nylund2018ten,howard2018variable,hipp2006local,hallquist2018mplusautomation}. For the 90 observations from our study, we examined how the observations from each active learning method were distributed across the identified profiles and compared student conceptual learning across profiles using Hedges' \textit{g}.

Four profiles of classroom activities emerged from our analysis: (1) ``Lecture," where the instructor lectures for most of the class time (blue dots in Fig.~\ref{fig:main2}b), (2) ``Clickers," where lecture is supplemented with group clicker questions (orange dots in Fig.~\ref{fig:main2}b), (3) ``Worksheets," where most of the class time is spent on group worksheets (yellow dots in Fig.~\ref{fig:main2}b), and (4) ``Other Groupwork," where most of the class time is spent on other types of groupwork activities (e.g., solving problems on whiteboards and completing laboratory activities; green dots in Fig.~\ref{fig:main2}b). For the 90 observations from our study, most Peer Instruction observations belong to the Clickers profile (with the rest belonging to the Lecture profile), ISLE observations are mostly split between the Lecture and Worksheets profiles, most Tutorials observations belong to the Worksheets profile, and most SCALE-UP observations belong to the Other Groupwork profile (Fig.~\ref{fig:main2}c). This means there are differences in the instructional practices used both within and across each of the four analyzed active learning methods. 

Assigning each course to the profile to which at least two of the three classroom observations belong, we find preliminary evidence that these instructional practices may be related to student conceptual learning (Fig.~\ref{fig:main2}d). The pooled effect size is larger for courses assigned to the Worksheets (Hedges' \textit{g} = 0.94 [0.60, 1.28]) and Other Groupwork (Hedges' \textit{g} = 0.95 [0.45, 1.45]) profiles than for courses assigned to the Lecture (Hedges' \textit{g} = 0.68 [0.33, 1.02]) and Clickers (Hedges' \textit{g} = 0.60 [0.34, 0.87]) profiles. However, a heterogeneity analysis indicates no significant variation in conceptual learning gains based on profile ($Q$ = 3.14, \textit{df} = 3, \textit{p} $=$ 0.37), possibly due to the Worksheets and Other Groupwork profiles having relatively small sample sizes. The differences in conceptual learning observed in Fig.~\ref{fig:main}, therefore, are likely related to differences in instructional activities \textit{and} other factors not measured here.

\section*{Discussion}

In response to calls for researchers to stop comparing active learning to traditional teaching methods and to instead compare different forms of active learning~\cite{dancy2024physics,wieman2014large}, we measured the relative impacts of four established active learning methods in introductory physics and astronomy: Peer Instruction, ISLE, Tutorials, and SCALE-UP. Using data from 28 institutions, we found significant increases in student conceptual learning in all four methods. 

We also found SCALE-UP to be associated with significantly larger conceptual learning gains than either Peer Instruction or ISLE. Surprisingly, these differences do not seem attributable to variation in the ways that students develop peer networks in each active learning method. Rather, the differences are at least partly related to variation in instructional practices. SCALE-UP courses spent the majority of class time on groupwork activities such as labs or problem solving at whiteboards, while instructors lectured and/or administered clicker questions for a large fraction of class time in many Peer Instruction and ISLE courses. This result has important implications for undergraduate physics instruction: while different active learning methods all coincide with meaningful student learning (moreso than traditional, lecture-oriented teaching methods), the extent and type of student-centered activities seems to be associated with different amounts of conceptual learning. We recommend that future research experiments collect similar data from a broader set of active learning physics courses (i.e., including courses using any form of active learning, not only a named active learning method) and compare conceptual learning gains across latent profiles of instructional practices (similar to our analysis in Fig.~\ref{fig:main2}d) to identify specific causal mechanisms of conceptual learning. 

Data from real classrooms contain many sources of variability, so it is worth noting several such sources which do not seem to alter this result. First, instructors decided which concept inventory was most relevant to their course, so several research-validated assessments were used. The Force Concept Inventory~\cite{hestenes1992force}, however, was the most commonly used instrument across all active learning methods (14 out of 26 courses) and the relative differences in conceptual learning between the methods remain the same when we only include courses that used this assessment (Supplementary Information Fig. 1).

We also studied courses from different disciplines (physics and astronomy), with different student populations (algebra-based and calculus-based courses), and from different types of institutions (PhD and non-PhD granting institutions and a range of research-intensive and undergraduate-focused institutions). It is possible that these variables are correlated with active learning method (e.g., if all of the SCALE-UP courses in our study were calculus-based) and/or student conceptual learning; however, a descriptive analysis of student learning across each of these factors indicates that this is not the case for our dataset (Supplementary Information Fig. 2). 

Variation in class size may also explain the relative benefits of active learning methods to conceptual learning. Prior research in other disciplines has found larger learning gains in small-enrollment active learning courses than large-enrollment active learning courses~\cite{murdoch2002active}. In this study, the ISLE, Tutorials, and SCALE-UP courses had similar class sizes, while Peer Instruction courses were larger, on average, than these three methods (Supplementary Information Table 1). Therefore, class size may help to explain why Peer Instruction exhibited the lowest effect size of the four methods, but it does not explain the differences in conceptual learning between ISLE and SCALE-UP.

Related to class size, the physical configuration of  active learning classrooms (e.g., tables set up for group work versus large lecture halls) may impact conceptual learning. In our study, however, a few of the Peer Instruction and ISLE courses took place entirely in classrooms with small tables for group work (similar to the SCALE-UP layout) rather than lecture halls. A prior study comparing two identically-taught introductory biology courses, one held in a SCALE-UP classroom and one held in a traditional lecture hall, also found that student learning did not vary across layouts~\cite{stoltzfus2016does}.  Our observed differences in conceptual learning gains between SCALE-UP, Peer Instruction, and ISLE, therefore, are likely not attributable to different physical classroom layouts.

Another significant finding is the within-method variation we observed in both classroom activities (Fig.~\ref{fig:main2}c) and student conceptual learning (Fig.~\ref{fig:main}). The variation in classroom activities among instructors using the same active learning method highlights an important theoretical implication of this study: the names of active learning methods in physics do not necessarily map onto instructional practices and thus may not be useful in differentiating active learning instruction. Researchers, therefore, should not solely rely on method names when making comparisons (e.g., of student outcomes) across instructional strategies. 

The within-method variation in student conceptual learning has several implications for future research. One possible explanation for higher conceptual learning gains in some courses is greater fidelity to the teaching method.  There is significant variation in instructor adaptation of established active learning methods~\cite{turpen2009not,wood2016characterizing} and without proper support, these adaptations may produce smaller conceptual learning gains than high-fidelity implementations~\cite{andrews2011active}. 
One route for future research, therefore, is to measure fidelity of implementation (e.g., by analyzing the nature and/or quality of classroom activities, such as worksheets) and relate this fidelity to student outcomes within and across methods. 

Relatedly, instructors' previous teaching experience may impact their choice and implementation of active learning methods. Newer instructors, for example, may be more likely to use Peer Instruction because it requires less overhaul of a course, whereas more experienced instructors may feel confident in implementing and adapting more complex methods such as SCALE-UP. Within each active learning method, more experienced instructors may exhibit larger student learning gains than newer instructors because they are more familiar with the method and their student population. We did not collect information about previous teaching experience from the instructors who participated in our study; however, we encourage future work to explore the relationships between instructor experience, active learning method, and student conceptual learning.

We note that we confined our observations to in-class activities and our learning measures to concept inventories. However, broader investigations to compare active learning methods might include out-of-class activities such as evidence-based transformations to homework, which can provide additional benefits to students in an active learning physics class~\cite{miller2021increasing}. A wider range of student outcomes may also be relevant to consider: for example, ISLE improves student engagement in scientific practices~\cite{etkina2010design}, but such practices may not translate to conceptual understanding~\cite{holmes2017value}. 

Finally, our choice to focus on four named methods was based on their robust research base and their broad popularity and availability through professional development. However, many physics instructors use a blend of different active learning methods -- including a number of potential participants who self-selected out of our study because they evaluated their teaching as not being a ``pure'' enough version of a named method. This underscores the need to develop vocabulary to describe physics learning environments that highlights key features and practices and does not rely on shorthand labels, whether that be ``traditional'' instruction, ``active learning,'' or specific named methods.

\textbf{Acknowledgments:} We thank all of the instructors and students who participated in our study. We also thank Karen Nylund-Gibson, Marsha Ing, and Adam Garber for their help with the Latent Profile Analysis. Jon Gaffney, Paula Heron, Natasha Holmes, and the Drexel Physics Education Research Network provided meaningful feedback on this manuscript. This material is based upon work supported by the National Science Foundation under Grant Nos. 2111128 (E. B. and A. T.) and 2224786 (E. B. and A. T.), and the Cotswold Foundation Postdoctoral Fellowship at Drexel University (M. S.).

\textbf{Author contributions:} A. Traxler and E. Brewe conceptualized and supervised the work; M. Sundstrom and J. Gambrell collected the data; M. Sundstrom, J. Gambrell, and C. Green applied the COPUS to the video observations; M. Sundstrom conducted the statistical analyses; all authors contributed to manuscript writing.

\textbf{Competing interests:} The authors declare no competing interests.

\clearpage

\section*{METHODS}

\section*{Participant recruitment and compensation}

We recruited instructors of introductory physics and astronomy courses (either algebra- or calculus-based) within the United States during the fall 2023, spring 2024, and fall 2024 semesters. We restricted data collection to first-semester physics (e.g., mechanics) and astronomy courses in order to have comparable measurements of peer network development across courses. In second-semester electromagnetism courses, for example, peer networks may have already developed in the previous mechanics course~\cite{sundstrom2022interactions}. 

To identify possible instructor participants, we first asked a grant advisory board member affiliated with each active learning method to provide a list of instructors who they knew implemented the method and sent emails to those instructors. We also posted advertisements within the American Physical Society and on Facebook. As data collection continued, we used snowball sampling, where we asked current participants if they knew other instructors using the same active learning method and reached out to them. After fall 2024, we could not identify any additional participants and concluded data collection.

31 instructors participated in the study, each with one course in one semester (i.e., no instructors participated in the study with multiple courses): 27 from physics and 4 from astronomy. Data were taken from distinct samples and did not include repeated measurements. All 31 instructors contributed data used in at least one of the three analyses presented in the main text. Instructors' data were only included in the analyses for which their data met the inclusion criteria (i.e., at least 40\% of and/or at least 30 enrolled students with matched responses to the concept inventory, at least 50\% pre- and post-semester response rates on the network survey, and three classroom video recordings). All instructors received \$1,000 as compensation.

\section*{Data analysis}

\subsubsection*{Concept inventories}

For each of the 26 courses with more than 40\% of and/or at least 30 enrolled students with matched responses on the concept inventory (i.e., completing both the pre- and post-semester assessment), we calculated an individual effect size---a standardized mean difference between students' pre- and post-semester concept inventory scores---using Hedges' \textit{g}~\cite{gurevitch1999statistical}. Hedges' \textit{g} is a form of Cohen's \textit{d}~\cite{cohen1992quantitative} that accounts for possible biases due to small sample sizes (e.g., small-enrollment physics courses). Hedges' \textit{g} is calculated as:
\begin{equation}
    g = \frac{\bar y_{post}-\bar y_{pre}}{s} \times (1 - \frac{3}{4n-9})
\end{equation}
where $\bar y_{post}$ is the mean student score on the post-concept inventory, $\bar y_{pre}$ is the mean student score on the pre-concept inventory, $s$ is the standard deviation of the distribution of differences in students' pre- and post-scores, and $n$ is the sum of the two sample sizes ($n = n_{post} + n_{pre}$). We performed these calculations using the ``cohen.d" function (with Hedges' correction) in the \textit{effsize} package in R (Version 4.4.2)~\cite{torchiano2016effsize}.

As in meta-analytic studies~\cite{freeman2014active}, we then combined the individual effect sizes to calculate a pooled effect size for each active learning method and determined whether each pooled effect size is significantly different from a null effect (i.e., an effect size of zero, indicating no difference in students' pre- and post-semester scores). We used a random effects model, which assumes the presence of both within-method variation (e.g., differences in student conceptual learning gains for courses using the same active learning method due to different instructors, different implementations, and/or different instructional contexts) and between-method variation (i.e., differences in student conceptual learning gains between active learning methods). The weight of each individual effect size, \textit{i}, is calculated as:
\begin{equation}
    w_i = \frac{1}{v_i + \tau^2}
\end{equation}
where $v_i$ is the within-method variance and $\tau^2$ is the between-method variance. The pooled effect size for each active learning method is calculated by taking the sum of each individual effect size multiplied by its weight and dividing that sum by the sum of all the weights:
\begin{equation}
    g_{pooled} = \frac{\sum_{i=1}^{k} w_i g_i}{\sum_{i=1}^{k} w_i}
\end{equation}
where \textit{k} is the number of courses using the active learning method and $g_i$ is the individual effect size for course \textit{i}. Therefore, $g_{pooled}$ can be interpreted as a weighted mean effect size. The 95\% confidence interval on each pooled effect size is given by $g_{pooled} \pm (1.96 \times s_{g_{pooled}})$, where $s_{g_{pooled}}$ is the standard deviation of the pooled effect size:
\begin{equation}
s_{g_{pooled}} = \sqrt{\frac{1}{\sum_{i=1}^{k} w_i}}.
\end{equation}
We calculated a $z$-value for each pooled effect size and compared it to the standard normal distribution to test if the pooled effect size is significantly different from zero (i.e., a null effect):
\begin{equation}
    z = \frac{g_{pooled}}{s_{g_{pooled}}}.
\end{equation}
This \textit{z}-value represents the difference between the pooled effect size and zero in units of standard deviation, or $\sigma$-differences. This analysis was performed using the ``rma" function in the \textit{metafor} package in R (Version 4.4.2)~\cite{viechtbauer2010conducting}.

We also determined whether the pooled effect sizes significantly differ between active learning methods by conducting a test of heterogeneity. This test quantifies whether the variation we see between the four pooled effect sizes is greater than chance (i.e., random noise). For this analysis, we first calculated the $Q$ statistic:
\begin{equation}
    Q = \sum_{j=1}^{4} w_j \times (g_{pooled, j} - \bar g_{pooled})^2
\end{equation}
where $w_j$ is the sum of weights within active learning method \textit{j}, $g_{pooled, j}$ is the pooled effect size for active learning method \textit{j}, and 
$\bar g_{pooled}$ is the mean pooled effect size across all four active learning methods. The degrees of freedom of $Q$ is equal to one less than the number of subgroups, in this case three (because there are four active learning methods). The statistical significance of $Q$, when compared to a chi-squared distribution, indicates whether variation in pooled effect sizes is due to random chance alone (if $Q$ is not significant) or instead due to some other source of variation (e.g., different active learning methods, if $Q$ is significant). 

As reported in the main text, $Q$ was significant in this study; therefore, we proceeded to conduct comparisons of pooled effect sizes between the different active learning methods. We used a mixed effects regression model with effect size as the dependent variable and active learning method as a moderator (Supplementary Information Table 4). This analysis was also performed using the ``rma" function in the \textit{metafor} package in R (Version 4.4.2).  

We checked that our results were not impacted by the different concept inventories used by each instructor by re-running the above analysis only including the 16 courses that used the Force Concept Inventory~\cite{hestenes1992force} (Supplementary Information Fig. 1). The relative differences between the methods replicated the results presented in the main
text, with the difference between SCALE-UP and Peer Instruction not retaining statistically significance because excluding large amounts of
our data set reduces statistical power. Only considering these courses using the Force Concept Inventory, we also observe that SCALE-UP exhibits significantly larger learning gains than Tutorials.

\subsubsection*{Peer networks}

As in prior work~\cite{commeford2021characterizing}, we created two undirected networks per course (one pre-semester and one post-semester), with nodes representing students and edges representing reported interactions between two students (regardless of which student reported the interaction on the survey). Treating the edges as undirected reduces possible impacts of missing data due to non-respondents because we assume an edge exists between two students even if only one of the students reported the interaction on the survey. Supplementary Information Figure 3 shows example networks from one course per active learning method. Networks for all courses can be obtained using the analysis scripts at Ref.~\cite{github2025}.

We aimed to measure network dynamics from the beginning to the end of the semester within each course. Most previous studies of social networks in physics education research use descriptive statistics to make claims about network patterns over time (e.g., Ref.~\cite{commeford2021characterizing}). These methods are limited, however, because they do not allow for hypothesis testing: they measure the effect of one variable at a time without accounting for other variables that may explain how networks change over time. Therefore, we applied temporal exponential random graph models (TERGMs) to the data to measure whether different network structures (e.g., triangles) are more (or less) likely to form in our observed post-semester networks than we would expect if the pre-semester networks evolved randomly. We used separable TERGMs in particular, as they model two distinct processes of network evolution: (1) formation---the tendency for new edges to form in the post-semester network that were not present in the pre-semester network, and (2) persistence---the tendency for edges formed in the pre-semester network to remain in the post-semester network~\cite{krivitsky2014separable}. In line with our research goal of identifying the types of network structures that form in different active learning methods, we only model the formation process in this study. 

We ran one TERGM for each of the 19 courses with sufficient response rates (at least 50\% on both the pre- and post-semester network surveys) using the following four predictor variables and the \textit{tergm} package in R (Version 4.4.2): edges, chains, triangles, and degree distribution (Supplementary Information Table 5). We included these particular variables because we expected them to vary between active learning methods. For example, we expected that chains would be more likely to form in Peer Instruction courses, as students typically sit in long rows of chairs in stadium style lecture halls, than in SCALE-UP courses, where students work in small groups at large tables~\cite{commeford2021characterizing}.

Before interpreting the model results, we ensured that all 19 models exhibited sufficient goodness-of-fit. We assessed model fit by simulating 100 networks using the resulting coefficient estimates and comparing the structural properties of these simulated networks to those of the observed networks~\cite{hunter2008goodness} (Supplementary Information Fig. 4).

\subsubsection*{Classroom observations}

We applied the full Classroom Observation Protocol for Undergraduate STEM (COPUS)~\cite{smith2013classroom} to each of the 90 classroom video recordings collected in our study. The COPUS documents the co-occurrence of 12 instructor activities and 13 student activities at two-minute time intervals.

Three of the authors contributed to the video coding. First, all three coders individually applied the COPUS to one full video observation from each of the four active learning methods. We then met to discuss disagreements and clarify inclusion and exclusion criteria for each COPUS activity. The three coders iteratively re-coded the videos and met for discussions until we reached greater than 80\% agreement (as measured with Cohen's Kappa~\cite{landis1977measurement}) for each of the four observations for each pair of coders (Supplementary Information Table 7). After reaching this level of agreement, we randomly assigned one coder to each of the remaining observations such that any resulting patterns among the observations were not attributable to systematic differences between coders. The first author (M. S.) coded 46 video observations, the second author (J. G.) coded 32 video observations, and the third author (C. G.) coded 12 video observations.

With these COPUS observations, we used Latent Profile Analysis (LPA) to identify representative profiles of classroom activities. LPA is a model-based, multi-variate approach to identifying unobserved latent subgroups (``profiles") in a dataset, where each group is characterized by a unique pattern of features (i.e., classroom activities). As LPA is a variation-based method, it requires a large sample size to achieve reliable results. While not fully agreed upon, research suggests a few hundred observations are necessary for LPA~\cite{nylund2018ten}. Therefore, we combined our 90 COPUS observations with 133 COPUS observations from two other studies (all 52 observations from Ref.~\cite{commeford2022characterizing} and a subset of 81 observations from Ref.~\cite{stains2018anatomy}). All of these observations come from introductory-level physics courses in the United States and were performed in the middle of the semester, indicating ``typical" class sessions. Any differences between these 223 observations, therefore, are likely attributable to variation in the instructional practices themselves and not any systemic differences across the study designs (e.g., the level or discipline of the course, the timing of the data collection, or the observation protocol used for analysis).

There are 25 different activities included in the full COPUS protocol; however, we only used a subset of eight of these activities in our LPA (Supplementary Information Table 8). This subset of activities is the same as that used in an LPA by Stains and colleagues~\cite{stains2018anatomy}, who found that these activities were not highly correlated with one another and meaningfully separated different profiles of COPUS observations. In another LPA of COPUS observations~\cite{commeford2022characterizing}, the authors included all 25 activities, but the resulting profiles were largely distinguished from one another by differences in these eight codes and not any others. Minimizing the number of variables (i.e., activities) included in the LPA, furthermore, assists with the reliability of the results because the model needs to estimate both a mean and a variance for each variable for each identified profile. The fewer the parameters being estimated, the more reliable each parameter estimate is.

As in prior work~\cite{stains2018anatomy,commeford2022characterizing}, the input variables for the LPA were the fraction of two-minute time intervals that each of the eight activities were present for each observation. We ran all four LPA model specifications for one- through six-profile solutions: (M1) equal variances in input variables across profiles and fixed covariances between input variables, (M2) freely estimated variances in input variables across profiles and fixed covariances between input variables, (M3) equal variances in input variables across profiles and freely estimated covariances between input variables, and (M4) freely estimated variances in input variables across profiles and freely estimated covariances between input variables. We estimated each model with 200 random sets of starting values, as recommended by prior work, to make sure that the model converged on a global rather than a local solution~\cite{hipp2006local}. We conducted the analysis using the \textit{MplusAutomation} package in R (Version 4.4.2)~\cite{hallquist2018mplusautomation}.

We evaluated model fit both within and across the four models. Based on current recommendations, we considered four information criteria to choose the optimal solution: Bayesian information criterion (BIC), sample size adjusted BIC (aBIC), consistent Akaike information criterion (CAIC), and approximate weight of evidence criterion (AWE; Supplementary Information Fig. 5). These information criteria pointed to the M1 4-class solution as the optimal model.

We then estimated each observation’s posterior probability of belonging to each profile and assigned it to the profile with the highest probability. We checked several model classification diagnostics to assess this model~\cite{masyn2013latent-key}. The entropy, a measure of overall precision of classification for the whole sample across all identified profiles, is 0.97 (values range from zero to one, with higher values indicating better precision), indicating that the model has reliable classification precision. The average posterior class probabilities for individuals assigned to each of the four profiles are all greater than 0.9, also indicating good classification. We also confirmed that this solution does not just separate observations from the three different data sources we combined in the analysis (Supplementary Information Table 10). 

Finally, we recognized that active learning method may not fully correlate with assigned instructional profile. Therefore, we assigned each instructor to the profile to which the majority of their three observations were assigned (only one course had three observations assigned to three different profiles and could not be included in this part of the analysis). We then calculated the pooled effect sizes of student conceptual learning for courses assigned to each profile using Hedges' \textit{g}. Similar to our previous analysis of student learning by active learning method, we conducted a test of heterogeneity to determine whether student learning varied by instructional profile.

\textbf{Data availability:}  All de-identified data used in this study can be found at \url{https://github.com/msundstrom33/ComparingActiveLearningMethods.git}.

\textbf{Code availability:} Analysis scripts for this study can be found at \url{https://github.com/msundstrom33/ComparingActiveLearningMethods.git}.

\textbf{Ethics statement:} This research was approved by  Drexel University's Institutional Review Board under Protocol No. 2203009130 and deemed exempt for board review as research within commonly accepted educational settings and involving educational surveys (Exemption Categories 1 and 2 of the United States Common Rule for Human Subjects Research). Informed consent was obtained from all human research participants.

\end{document}